\documentclass[prb,aps,twocolumn,showpacs]{revtex4}
\usepackage{bm}
\usepackage{amssymb}
\usepackage{graphicx}

\begin{document}

\title{Decay of quasiparticles in quantum spin liquids}

\author{M. E. Zhitomirsky}

\affiliation{
Commissariat \`a l'Energie Atomique, DSM/DRFMC/SPSMS, 38054 Grenoble,
France}

\date{\today}

\begin{abstract}
Magnetic excitations are studied in gapped
quantum spin systems, for which spontaneous two-magnon decays
are allowed by symmetry.
Interaction between one- and two-particle
states acquires nonanalytic frequency and momentum dependence near
the boundary of two-magnon continuum. This leads to 
a termination point of single-particle branch for one-dimensional
systems and to strong suppression of quasiparticle weight in 
two dimensions. The momentum dependence of the decay rate is
calculated in arbitrary dimensions and effect of external
magnetic field is discussed. 
\end{abstract}

\maketitle

\begin{figure}[b]
\centerline{
\includegraphics[width=0.9\columnwidth]{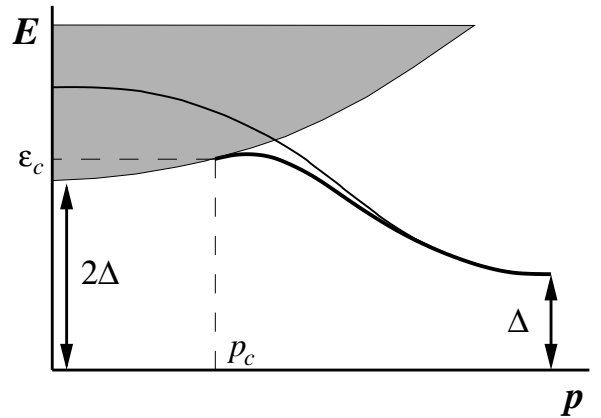}}
\caption{Schematic structure of the energy spectrum of a quantum magnet
with spin-liquid ground state.
Thin solid line is a bare spectrum $\varepsilon({\bf p})$, full
solid line is a renormalized single-particle branch, shaded region
is two-particle continuum. The decay instability threshold 
is denoted by $p_c$.
}
\label{crossing}
\end{figure}

A number of magnetic materials with quantum disordered, or spin-liquid,
ground states have been discovered in the past two decades.
The well-known examples include the spin-Peierls compound CuGeO$_3$,
\cite{cugeo} dimer systems Cs$_3$Cr$_2$Br$_9$ \cite{cscrbr} and  
TlCuCl$_3$, \cite{tlcucl} 
integer-spin antiferromagnetic chains, \cite{Hchain} and 
many others. Common property of all these systems is
presence of a spin gap in the excitation spectrum, which separates
a singlet, $S=0$ ground state from $S=1$ quasiparticles.
The low-energy triplet of magnons can be split 
due to intrinsic anisotropies or under applied magnetic field.
Recent neutron experiments\cite{phcc,ipa} 
on two organic spin-gap materials PHCC and 
IPA-CuCl$_3$ have revealed drastic transformation
of triplet quasiparticles undergoing at high energies. Upon
entering two-magnon continuum, see Fig.~\ref{crossing}, 
spontaneous ($T=0$) decay of 
a magnon into a pair of quasiparticles becomes possible,
which leads to a rapid decrease in the quasiparticle life-time in the 
former case\cite{phcc}  and complete disappearance of the single-particle
branch in the latter system.\cite{ipa} 

Quasiparticle instability 
is well documented for another type of quantum 
liquid---superfluid $^4$He. Predicted
by Pitaevskii nearly fifty years ago, \cite{pitaevskii,LL,zawad} 
this instability was later confirmed by 
neutron scattering measurements. \cite{smith,fak}  
In liquid helium, interaction between one-
and two-particle states is enhanced in the vicinity of a decay
threshold by a large density of roton states and  
leads to avoided crossing: the single-particle branch
flattens at energies below twice the roton energy and ceases
to exist completely above that energy scale.

\begin{figure}[t]
\centerline{
\includegraphics[width=0.9\columnwidth]{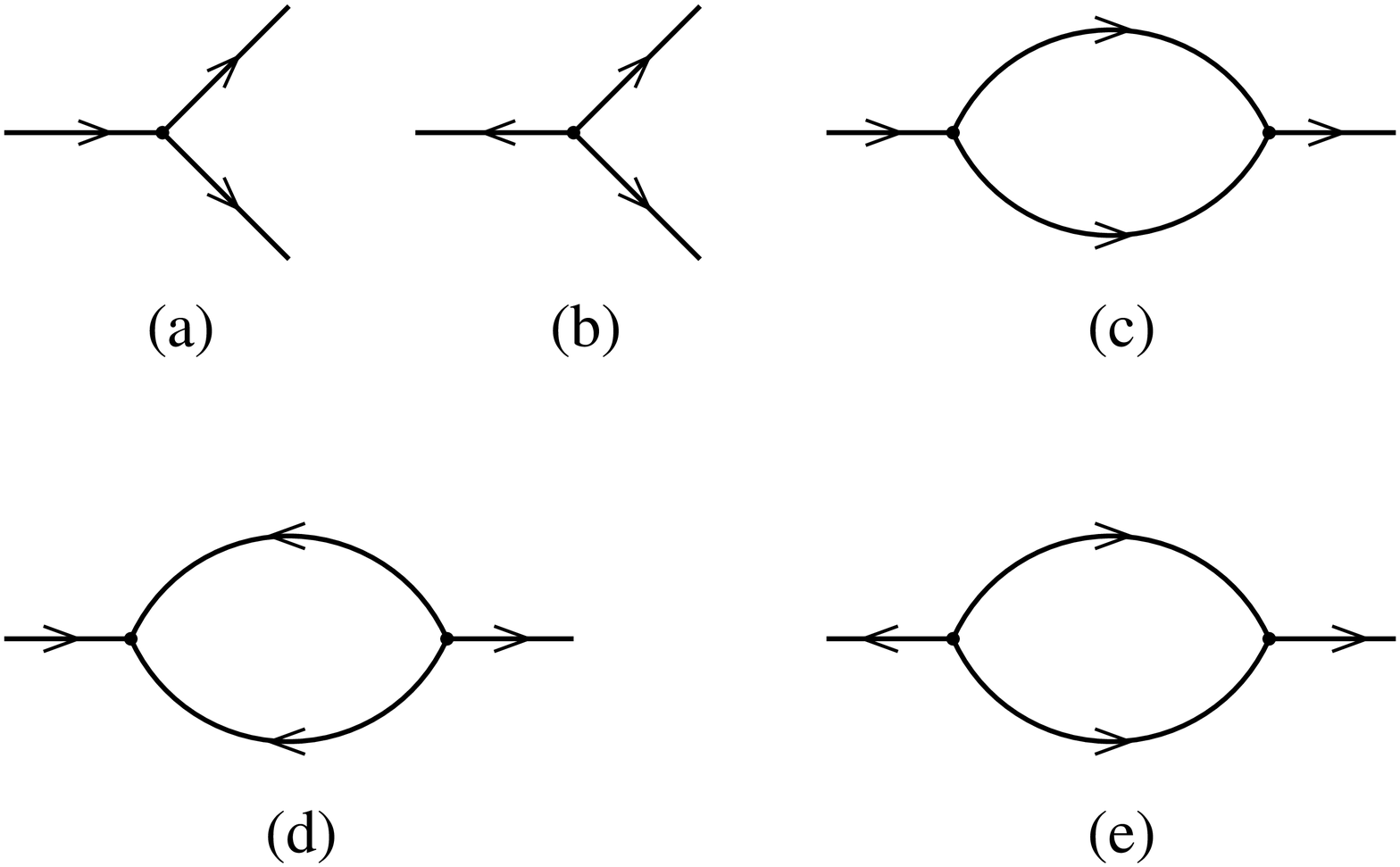}}
\caption{Two cubic vertices of the decay (a) and the source (b) types
and their contributions to the normal (c) and (d) and anomalous (e)
self-energies.}
\label{diagrams}
\end{figure}

The aim of present article is to investigate similar effects
in quantum spin liquids. We consider an isotropic spin system with
a quantum disordered ground state, which is separated by 
a finite gap $\Delta$ from low-energy spin-1 excitations. 
A bare dispersion of propagating triplets
is given by $\varepsilon({\bf p})$.
Bosonic operators $t_{{\bf p}\alpha}$ and $t^\dagger_{{\bf p}\alpha}$  
destroy and create a quasiparticle with 
momentum $\bf p$ in one of the three polarizations $\alpha=x,y,z$. 
Interaction between one- and two-particle states is, generally, described by
two types of cubic vertices shown in Fig.~\ref{diagrams}a,b.
In superfluid $^4$He presence of such particle non-conserving processes 
is determined by a Bose condensate, which absorbs or emits 
an extra particle. \cite{LL,fetter}
In quantum spin systems with singlet ground states one finds a 
more diverse situation. Symmetry of a majority 
of low-dimensional dimer systems, like, for example,  a dimerized
chain,\cite{harris} an asymmetric ladder and a planar array 
of dimers,\cite{sachdev} allows presence of cubic vertices.
These should be contrasted with a symmetric ladder, \cite{goplan}
where one- and two-magnon states belong to sectors with different parity 
under permutation of two legs and, consequently, do not interact.
There is also a suggestion that cubic vertices exists in 
a Heisenberg antiferromagnetic spin-1 chain beyond the nonlinear sigma-model 
description. \cite{kolezhuk}
In the following, we shall assume presence of interaction
between one- and two-particles states and consider
its consequences for the dynamic properties of a spin liquid.

In Heisenberg magnets, rotational symmetry in the spin space
fixes uniquely the tensor structure of cubic vertices.
Specifically, the decay vertex, Fig.~\ref{diagrams}a, has the 
following form:
\begin{equation}
\hat{V}_3 = \frac{1}{2!} \sum_{{\bf k},{\bf q}} \Gamma({\bf k},{\bf q})
\,\epsilon^{\alpha\beta\gamma}\: t^\dagger_{{\bf k}\alpha }
t^\dagger_{{\bf q}\beta }
 t^{_{}}_{{\bf k}+{\bf q}\gamma } \ .
\label{decayV}
\end{equation}
Conservation of the total spin during decay process
requires that two created spin-1 quasiparticles must form an
$S=1$ state, which is imposed by the antisymmetric tensor
$\epsilon^{\alpha\beta\gamma}$. The vertex is, consequently,
antisymmetric under permutation of two momenta 
$\Gamma({\bf k},{\bf q})=-\Gamma({\bf q},{\bf k})$.
Similar consideration applies to the source-type vertex, 
Fig.~\ref{diagrams}b, which is antisymmetric under permutation
of any two of three outgoing lines.

Important kinematic property of the energy spectrum
$\varepsilon({\bf p})$ is a type of instability
at a decay threshold momentum ${\bf p}_c$, beyond
which the energy conservation
\begin{equation}
\varepsilon({\bf p})=\varepsilon({\bf q})+\varepsilon({\bf p}-{\bf q})
\label{decay}
\end{equation}
is satisfied and two-magnon decays become possible.
Extremum condition imposed on the right hand side of Eq.~(\ref{decay})
yields that two quasiparticles at the bottom of continuum 
always have equal velocities ${\bf v}_{\bf Q} = 
{\bf v}_{{\bf p}-{\bf Q}}$, where  
${\bf v}_{\bf k}= \nabla_{\bf k}\,\varepsilon({\bf k})$. 
Then, the simplest possibility is 
that both momenta are also equal
with ${\bf Q} = \frac{1}{2}({\bf p}+{\bf G})$, where
${\bf G}$ is a reciprocal lattice vector.
We have verified the above assertion for
several model dispersion curves \cite{uhrig,oitmaa} 
as well as for a few experimental fits. \cite{phcc,ipa,csnicl3}
The common feature of all analyzed cases is presence of two regimes:
for a small total momentum $\bf p$
of a magnon pair the minimum energy corresponds to equal
momenta of two quasiparticles, whereas for $\bf p$ 
near the Brillouin zone boundary 
two particles of the lowest-energy pair have different momenta.
Such a bifurcation is
closely related to appearance of bound pairs of magnons 
at large momenta for one-dimensional
spin systems. \cite{harris,sushkov}
Kinematic instability at the decay threshold in all considered cases
corresponds, however, 
to a decay into a pair of quasiparticles with equal momenta.
In the following we shall focus on such type of a model-independent 
instability.

Let us consider the second-order contribution
to the normal self-energy 
from the decay processes shown in Fig.~\ref{diagrams}c.
Standard calculation yields
\begin{equation}
\Sigma_{11}(\omega,{\bf p}) = 2 \int \frac{d^D q}{(2\pi)^D}
\frac{\Gamma({\bf q}, {\bf p}-{\bf q})^2}
{\omega - \varepsilon({\bf q}) - 
\varepsilon({\bf p}-{\bf q}) + i0} \ ,
\label{sigma11}
\end{equation}
where $D$ is the number of dimensions. We shall be interested
in the behavior of energy spectrum at small
$\Delta{\bf p}={\bf p}-{\bf p}_c$ and 
$\Delta\omega=\omega-\varepsilon_c$  
and expand accordingly all functions under integral in 
$\Delta{\bf q}= {\bf q}-{\bf Q}$, where ${\bf Q} = \frac{1}{2}
({\bf p}+{\bf G})$.   The main difference with Pitaevskii's analysis
\cite{LL} is that the antisymmetric vertex vanishes
at the decay threshold such that 
\begin{equation}
\Gamma({\bf Q}+\Delta{\bf q},{\bf Q}-\Delta{\bf q})\approx 2\Delta {\bf q} 
\cdot\nabla_{\bf k}\Gamma({\bf k},{\bf q})|_{{\bf k},{\bf q}={\bf Q}} \ .
\end{equation}
Including all dimensional constants in the vertex 
$g_3\sim |\nabla_{\bf k}\Gamma({\bf k},{\bf q})|$
and performing angular integration for $D>1$ 
we obtain 
\begin{equation}
\Sigma_{11}(\omega,{\bf p})\simeq - g_3^2 \int_0^\Lambda \frac{q^{D+1}\,dq}
{q^2 + {\bf v}_2\Delta{\bf p}-\Delta\omega -i0} \ ,
\label{sigma}
\end{equation}
where
${\bf v}_2=\nabla_{\bf k}\,\varepsilon({\bf k})|_{{\bf k}={\bf Q}}$
and $\Lambda$ is a lattice cut-off.
The self-energy remains finite as $\Delta\omega,|\Delta{\bf p}|
\rightarrow 0$, though it contains an important 
nonanalytic contribution $\tilde{\Sigma}(\omega,{\bf p})$.
We shall absorb a regular part of $\Sigma_{11}(\omega,{\bf p})$
into $\varepsilon({\bf p})$ and determine 
a new corrected spectrum from the Dyson's equation:
$G^{-1}(\omega,{\bf p}) = \omega -\varepsilon({\bf p}) - 
\tilde\Sigma(\omega,{\bf p})=0$.

Two other second-order diagrams constructed from cubic vertices
are presented in Fig.~\ref{diagrams}d,e. Correction to the normal
self-energy shown in Fig.~\ref{diagrams}d is an analytic 
function of frequency and momentum and can be  
absorbed into $\varepsilon({\bf p})$. 
The anomalous self-energy $\Sigma_{12}(\omega,{\bf p})$,
Fig.~\ref{diagrams}e, has the same dependence  
on $\omega$ and $\bf p$ as the one-loop diagram (\ref{sigma}).
Using the Belyaev's expression for the normal Green's 
function of a Bose gas: \cite{LL,fetter}
\begin{eqnarray}
G^{-1}(\omega,{\bf p}) & = & \omega-\varepsilon({\bf p}) - 
\Sigma_{11}(\omega,{\bf p})  \nonumber \\
&& \mbox{} + \frac{\Sigma_{12}(\omega,{\bf p})\Sigma_{21}(\omega,{\bf p})}
{\omega+\varepsilon(-{\bf p}) + \Sigma_{11}(-\omega,-{\bf p})} \ ,
\end{eqnarray}
one can conclude that inclusion of the anomalous self-energy
simply modifies a coefficient in front of the nonanalytic 
contribution (\ref{sigma}) without changing its functional
dependence.
Similar result applies to the vertex correction from
multiple two-particle scattering processes.
Since a one-loop diagram is finite for $\omega=\varepsilon_c$  
${\bf p}={\bf p}_c$, the renormalized vertex $\tilde{g}_3$
also remains finite and the vertex correction amounts to 
replacement $g_3^2 \rightarrow g_3\tilde{g}_3$ in front
of the integral in Eq.~({\ref{sigma}). 
Thus, the following analysis based on analytic properties
of the one-particle Green's function is not restricted to the lowest-order
perturbation theory used in derivation of Eq.~(\ref{sigma11}) and is
quite general.

\paragraph*{One-dimensional gapped spin systems:}
Let us begin with a stable region, where 
spontaneous decays are forbidden:
$\Delta\omega<0$, $\Delta p>0$, such that $(v_2\Delta p-\Delta\omega)>0$.
Separating the nonanalytic part in the integral (\ref{sigma}) and expanding
$\varepsilon(p) \approx \varepsilon_c + v_1\Delta p$ 
we obtain for the inverse Green's function
\begin{equation}
G^{-1}(\omega,p) = \Delta\omega - v_1\Delta p - 
\lambda \sqrt{v_2\Delta p -\Delta\omega} \ ,
\label{G1d}
\end{equation}
with $\lambda = \pi g_3^2/2$. The one-particle Green's function 
with such a dependence
on $\omega$ and $p$ has been briefly discussed in Ref.~\onlinecite{LL}.
In the following we give more details relevant for the
case of Eq.~(\ref{G1d}).

Condition for a pole  $G^{-1}(\omega,p)=0$ is transformed to 
quadratic equation, which yields 
\begin{equation}
\bar{\varepsilon}(p)=\varepsilon_c + v_1\Delta p - \frac{\lambda^2}{2}
+ \sqrt{\frac{\lambda^4}{4} + \lambda^2 (v_2-v_1)\Delta p} \ .
\label{Eren}
\end{equation}
Away from the crossing point for $\Delta p\gg\lambda^2/(v_2-v_1)$,
slope of the single-particle branch 
coincides with the bare magnon velocity 
$d\bar{\varepsilon}(p)/dp\approx v_1<0$, whereas 
for $\Delta p\rightarrow 0$ slope changes sign to
$d\bar{\varepsilon}(p)/dp = v_2>0$.  
In the crossover region $\Delta p\sim\lambda^2/(v_2-v_1)$
the quasiparticle weight $Z$,
\begin{equation}
Z^{-1}\!=\left.\!\frac{\partial G^{-1}}{\partial \omega}\right|_
{\omega=\bar{\varepsilon}}\!\! = 1 + \frac{\lambda^2 +
\sqrt{\lambda^4+4\lambda^2(v_2-v_1)\Delta p}}
{4(v_2-v_1)\Delta p}\,,
\end{equation}
is continuously suppressed and vanishes, once the single-particle
branch touches the two-magnon continuum. 

On the opposite side 
of the crossing point, $\Delta p<0$, $\Delta\omega>0$, and 
$(\Delta\omega -v_2\Delta p)>0$, 
the nonanalytic part of the self-energy becomes purely imaginary
$\tilde{\Sigma} = -i\lambda\sqrt{\Delta\omega-v_2\Delta p}$.
Formally, condition $G^{-1}(\omega,p)=0$ yields after 
transformation to quadratic equation the same
solution as in the stable region.
For small negative $\Delta p$ the root (\ref{Eren}) 
is real and satisfies 
$\Delta\omega-v_1\Delta p<0$ and $\Delta\omega-v_2\Delta p<0$.
The latter of the above two relations contradicts the assumption 
made in the beginning of the paragraph for the decay region,
whereas the former is inconsistent with zero of $G^{-1}(\omega,p)$
in the region, where $v_2\Delta p-\Delta\omega>0$, see Eq.~(\ref{G1d}).
Thus, inside the continuum, a physical pole of the 
one-particle Green's function reappears only at a finite
distance from the crossing point, where ${\rm Re}\,\{\Delta\omega\}>
v_2\Delta p$  or
\begin{equation}
\Delta p < - \frac{\lambda^2}{2(v_2-v_1)} \ .
\end{equation}
Disappearance of  single-magnon excitations inside the continuum
is quite similar to the termination point in the energy spectrum
of superfluid $^4$He. \cite{pitaevskii,LL,zawad,smith,fak}
In both cases the  single-particle branch approaches
tangentially the boundary of the two-particle continuum.

Further away from the lower edge of the continuum in the 
region, where $\bar{\varepsilon}(p)\approx \varepsilon_c+v_1\Delta p$,
one can find for the inverse quasiparticle life-time:
\begin{equation}
\gamma_p \simeq 2\lambda \sqrt{(v_2-v_1)|\Delta p|} \ .
\label{lifetime1}
\end{equation}
Such a square-root dependence of the decay rate 
follows also from a simple golden rule consideration. \cite{kolezhuk}
We have seen, however, that the full dependence
of the self-energy on frequency and momentum must be kept
in the vicinity of the decay threshold.
Besides, if the cubic vertices have no special smallness,
{\it i.e.}, $\lambda$ is of the order of the bandwidth of
$\varepsilon(p)$, the perturbative regime, where
Eq.~(\ref{lifetime1}) applies, is never realized.
Such a situation apparently occurs
in the quasi one-dimensional
gapped spin system IPA-CuCl$_3$,
where no trace of single-magnon excitations
has been observed inside 
the continuum.\cite{ipa}

\paragraph*{Two-dimensional gapped spin systems:}
Below decay threshold in the stable region, where
$({\bf v}_2\Delta{\bf p}-\Delta\omega)>0$, the nonanalytic part
of the self-energy becomes
\begin{equation}
\tilde{\Sigma}(\omega,{\bf p}) = \lambda ({\bf v}_2\Delta{\bf p}
 - \Delta\omega) \ln\frac{R}{{\bf v}_2\Delta{\bf p}-\Delta\omega} \ ,
\end{equation}
where $\lambda=g_3^2/2$ and $R$ is an energy cut-off.
Though, a pole of the Green's function cannot be found analytically,
it is easy to verify that solution of $G^{-1}(\omega,{\bf p})=0$
exists all the way down to the crossing point. The quasiparticle
branch touches tangentially the boundary of 
the continuum $\nabla_{\bf p}\bar{\varepsilon}({\bf p})={\bf v}_2$
for ${\bf p}={\bf p}_c$, whereas the quasiparticle weight
is gradually diminished and vanishes at the boundary.

Above the decay threshold [$({\bf v}_2\Delta{\bf p}-\Delta\omega)<0$]
the self-energy acquires imaginary part:
\begin{equation}
\tilde{\Sigma}(\omega,{\bf p}) = -\lambda (\Delta\omega -
{\bf v}_2\Delta{\bf p}) \left(\ln\frac{R}{\Delta\omega - 
{\bf v}_2\Delta{\bf p}} + i\pi\right)  . 
\end{equation}
In the perturbative regime, where $\bar{\varepsilon}({\bf p})
\approx \varepsilon_c + {\bf v}_1\Delta{\bf p}$, the decay rate
of magnons 
grows linearly inside the continuum:
\begin{equation}
\gamma_p \simeq 2\pi\lambda ({\bf v}_1-{\bf v}_2)\Delta {\bf p} \ .
\label{lifetime2}
\end{equation}
In the close vicinity of the crossing point we write instead 
$\bar{\varepsilon}({\bf p})\approx {\bf v}_2\Delta{\bf p}+a$ 
with ${\rm Re}\,a>0$ and 
transform equation on the pole of the Green's function to
\begin{equation}
a \left[ \ln\frac{R}{a} + i\pi\right] =   \frac{
({\bf v}_1-{\bf v}_2)\Delta {\bf p}}{\lambda}=b > 0 \ .
\end{equation}
With logarithmic accuracy the
solution of the above equation  is
\begin{equation}
{\rm Re}\,a = \frac{b}{\sqrt{\ln^2R/b+\pi^2}} \ , \ \
{\rm Im}\,a = \frac{-\pi b}{\ln^2R/b+\pi^2} \ .
\end{equation}
Single-particle branch of a two-dimensional gapped 
spin system can be, therefore, continued inside the decay region.  
However, the quasiparticle weight is suppressed
in a range of momenta near the crossing point:
\begin{equation}
|({\bf v}_1-{\bf v}_2)\Delta{\bf p}|\simeq R e^{-1/\lambda} \ .
\end{equation}
Suppression of the quasiparticle peak has been experimentally
observed in two-dimensional
quantum disordered antiferromagnet PHCC. \cite{phcc}

\paragraph*{Three-dimensional gapped spin systems:}
The nonanalytic part of the self-energy is given in this case by
\begin{eqnarray}
&& \tilde{\Sigma}(\omega,{\bf p}) = 
-\lambda({\bf v}_2\Delta{\bf p}-\Delta\omega)^{3/2}, 
\ \ \ \ \ \ \ \
\Delta\omega <{\bf v}_2\Delta{\bf p}\ , 
\nonumber \\
&& \tilde{\Sigma}(\omega,{\bf p}) = 
-i\lambda(\Delta\omega- {\bf v}_2\Delta{\bf p})^{3/2},
\ \ \ \ \ \ \  
\Delta\omega >{\bf v}_2\Delta{\bf p}\ , \nonumber
\end{eqnarray}
where $\lambda = \pi g_3^2/2$.
Such a weakly nonanalytic term can be treated perturbatively
near the crossing point.
In particular, the quasiparticle weight is not suppressed as
${\bf p}\rightarrow {\bf p}_c$ and $\bar{\varepsilon}({\bf p})$
does not change its slope at the crossing.
The magnon decay rate in three-dimensional case is 
\begin{equation}
\gamma_p \simeq 2\lambda [({\bf v}_1-{\bf v}_2)\Delta {\bf p}]^{3/2} \ .
\label{lifetime3}
\end{equation}

An experimentally relevant question is how an external 
magnetic field affects the decay processes. If the Zeeman
energy is smaller than the spin gap $g\mu_BH<\Delta$,
the role of an applied field reduces
to splitting a triplet of low-energy excitations.
This can, in principle, lead to breaking the resonance condition
(\ref{decay}).
To study such a possibility one should transform from the vector basis
$t_{\bf p\alpha}$ to states with a definite spin projection
on the field direction: 
\begin{equation}
t_{{\bf p}0} =  t_{{\bf p}z}\ , \ \ \ \ 
t_{\bf p\pm} = \mp \frac{1}{\sqrt{2}}\left(t_{{\bf p}x}\mp it_{{\bf p}y}
\right) .
\end{equation}
The decay vertex (\ref{decayV}) taken in the new basis has 
the following form
\begin{eqnarray}
&& i \epsilon^{\alpha\beta\gamma} t^\dagger_{{\bf p}\alpha }
t^\dagger_{{\bf q}\beta } t^{_{}}_{{\bf k}\gamma } =
\bigl(t^\dagger_{{\bf p}+}t^\dagger_{{\bf q}-} -
t^\dagger_{{\bf p}+}t^\dagger_{{\bf q}-} \bigr)t^{_{}}_{{\bf k}z} 
 \\
&&  \ \ \mbox{} 
+ \bigl(t^\dagger_{{\bf p}+}t^\dagger_{{\bf q}z} -
t^\dagger_{{\bf p}z}t^\dagger_{{\bf q}+} \bigr)t^{_{}}_{{\bf k}+} 
+ \bigl(t^\dagger_{{\bf p}z}t^\dagger_{{\bf q}-} -
t^\dagger_{{\bf p}-}t^\dagger_{{\bf q}z} \bigr)t^{_{}}_{{\bf k}-}.
\nonumber 
\end{eqnarray}
In all decay channels 
a destroyed one-particle and a created two-particle states
experience the same energy shifts.
Therefore, the decay threshold momentum ${\bf p}_c$ 
does not change for all three splitted branches of the
spin-1 excitations.

Once the Zeeman energy exceeds the triplet gap, $H>H_c=\Delta/g\mu_B$,
a Bose condensation of magnons takes place in $D>1$. \cite{nikuni}
The ground state acquires a nonzero magnetization and 
a long-range order of transverse spin components.
Such a canted antiferromagnetic structure opens an additional
channel for spontaneous decays of low-energy excitations in 
the gapless branch, similar to the prediction made for
ordered antiferromagnets in the vicinity of the saturation
field. \cite{mzh} 

Intrinsic magnetic anisotropies, which become important
in materials with spins $S\geq 1$, can also affect 
the decay processes. Anisotropy changes differently
dispersion of triplet excitations and modifies 
the resonance condition (\ref{decay}). 
The decay rate of spin-1 excitations depends, then, 
on a polarization, as was experimentally
observed in bond-alternating quasi-one-dimensional
antiferromagnet NTENP. \cite{hagiwara}   
The tensor structure of the decay vertex (\ref{decayV}) can be also
modified due to the absence of spin-rotational symmetry.
Dynamic properties of spin liquids with
different types of anisotropies deserve further 
theoretical investigations.

In conclusion, by studying analytic properties
of the one-particle Green's function near the decay threshold
we have found that crossing of a single-particle branch 
into two-magnon continuum is described solely by 
a growing line-width
of magnons only for three-dimensional quantum spin liquids.
In two dimensions there is, in addition, strong suppression of
a one-magnon peak near the crossing point, whereas
in one dimension a single-magnon branch terminates
at the continuum boundary.

I thank A. Abanov and I. Zaliznyak for helpful
discussions.
The hospitality of
the Condensed Matter Theory Institute of the Brookhaven
National Laboratory, where the  
present work has been started, is gratefully acknowledged.


\begin{thebibliography}{99}

\bibitem{cugeo}
M. Hase, I. Terasaki, and K. Uchinokura,
Phys. Rev. Lett. {\bf 70}, 3651 (1993).

\bibitem{cscrbr}
B. Leuenberger, H. U. G\"udel, R. Feile, and J. K. Kjems,
Phys. Rev. B {\bf 28}, R5368 (1983).

\bibitem{tlcucl}
A. Oosawa, M. Ishi, and H. Tanaka, 
J. Phys.: Condens. Matter {\bf 11}, 265 (1999).

\bibitem{Hchain}
L. P. Regnault, I. Zaliznyak, J. P. Renard, and C. Vettier, 
Phys. Rev. B {\bf 50}, 9174 (1994).

\bibitem{phcc}
M. B. Stone, I. A. Zaliznyak, T. Hong, C. L. Broholm, and D. H. Reich,
{\tt e-print:\ cond-mat/0511266}.

\bibitem{ipa}
T. Masuda, A. Zheludev, H. Manaka, L.-P. Regnault, J.-H. Chung, and Y. Qiu,
{\tt e-print:\ cond-mat/0511143}.

\bibitem{pitaevskii}
L. P. Pitaevskii, Zh. \'Eksp. Teor. Fiz. {\bf 36}, 1168 (1959)
[Sov. Phys. JETP {\bf 9}, 830 (1959)].

\bibitem{LL}
E. M. Lifshits and L. P. Pitaevskii, {\it Statistical Physics II}
(Pergamon, Oxford, 1980).

\bibitem{zawad}
A. Zawadowski, J. Ruvalds, and J. Solana, 
Phys. Rev. A {\bf 5}, 399 (1972).

\bibitem{smith}
A. J. Smith, R. A. Cowley, A. D. B. Woods, W. G. Stirling,
and P. Martel, J. Phys. C {\bf 10}, 543 (1977).

\bibitem{fak}
B. F\aa k and J. Bossy, J. Low Temp. Phys. {\bf 112}, 1 (1998).

\bibitem{fetter}
A. L. Fetter and J. D. Walecka,
{\it Quantum theory of many-particle systems} (McGraw-Hill, New York, 1971).

\bibitem{harris}
A. B. Harris,
Phys. Rev. B {\bf 7}, 3166 (1973).

\bibitem{sachdev}
S. Sachdev and R. N. Bhatt,
Phys. Rev. B {\bf 41}, 9323 (1990).

\bibitem{goplan}
S. Gopalan, T. M. Rice, and M. Sigrist,
Phys. Rev. B {\bf 49}, 8901 (1994).

\bibitem{kolezhuk}
A.~Kolezhuk and S.~Sachdev, 
{\tt e-print:\ cond-mat/ 0511353}.

\bibitem{uhrig}
C. Knetter and G. S. Uhrig, 
Eur. Phys. J. B {\bf 13}, 209 (2000).

\bibitem{oitmaa}
J. Oitmaa, R. R. P. Singh, and W. Zheng,
Phys. Rev. B {\bf 54}, 1009 (1996).

\bibitem{csnicl3}
I. A. Zaliznyak, S.-H. Lee, and S. V. Petrov,
Phys. Rev. Lett. {\bf 87}, 017202 (2001).

\bibitem{sushkov}
O. P. Sushkov and V. N. Kotov, 
Phys. Rev. Lett. {\bf 81}, 1941 (1998).

\bibitem{nikuni}
T. Nikuni, M. Oshikawa, A. Oosawa, and H. Tanaka 
Phys. Rev. Lett. {\bf 84}, 5868 (2000).

\bibitem{mzh}
M. E. Zhitomirsky and A. L. Chernyshev,
Phys. Rev. Lett. {\bf 82}, 4536 (1999).

\bibitem{hagiwara}
M. Hagiwara, L. P. Regnault, A. Zheludev, A. Stunault,  N. Metoki, T.
Suzuki, S. Suga, K. Kakurai, Y. Koike, P. Vorderwisch, and J. H. Chung,
Phys. Rev. Lett  {\bf 94}, 177202 (2005).


\end{thebibliography}
\end{document}